\documentclass[sigconf,natbib=true]{acmart}
\usepackage{subfigure}
\usepackage{soul}
\usepackage{multirow}
\theoremstyle{plain}

\newtheorem{myTheo}{Theorem}
\usepackage{makecell}
\makeatletter
\newif\if@restonecol
\makeatother

\usepackage[ruled,vlined,linesnumbered]{algorithm2e}
\let\oldnl\nl
\newcommand{\nonl}{\renewcommand{\nl}{\let\nl\oldnl}}
\AtBeginDocument{%
  \providecommand\BibTeX{{%
    \normalfont B\kern-0.5em{\scshape i\kern-0.25em b}\kern-0.8em\TeX}}}

\setcopyright{acmcopyright}
\copyrightyear{2023}
\acmYear{2023}
\setcopyright{acmlicensed}\acmConference[SIGIR '23]{Proceedings of the 46th International ACM SIGIR Conference on Research and Development in Information Retrieval}{July 23--27, 2023}{Taipei, Taiwan}
\acmBooktitle{Proceedings of the 46th International ACM SIGIR Conference on Research and Development in Information Retrieval (SIGIR '23), July 23--27, 2023, Taipei, Taiwan}
\acmPrice{15.00}
\acmDOI{10.1145/3539618.3591948}
\acmISBN{978-1-4503-9408-6/23/07}

\title{Always Strengthen Your Strengths: A Drift-Aware Incremental Learning Framework for CTR Prediction}

\author{Congcong Liu*, Fei Teng*, Xiwei Zhao, Zhangang Lin, Jinghe Hu, Jingping Shao }
\affiliation{%
  \institution{JD.com}
  \country{Beijing, China}
  }

\email{ {liucongcong25,tengfei49,zhaoxiwei,linzhangang,hujinghe,shaojingping}@jd.com }
\email{tf20@mails.tsinghua.edu.cn}

\renewcommand{\shortauthors}{Liu and Teng, et al.}
\begin{document}

\begin{abstract}
Click-through rate (CTR) prediction is of great importance in recommendation systems and online advertising platforms. 
When served in industrial scenarios, the user-generated data observed by the CTR model typically arrives as a stream. 
Streaming data has the characteristic that the underlying distribution drifts over time and may recur.
This can lead to catastrophic forgetting if the model simply adapts to new data distribution all the time.
Also, it's inefficient to relearn distribution that has been occurred.
Due to memory constraints and diversity of data distributions in large-scale industrial applications, conventional strategies for catastrophic forgetting such as replay, parameter isolation, and knowledge distillation are difficult to be deployed.
In this work, we design a novel drift-aware incremental learning framework based on ensemble learning to address catastrophic forgetting in CTR prediction. 
With explicit error-based drift detection on streaming data, the framework further strengthens well-adapted ensembles and freezes ensembles that do not match the input distribution avoiding catastrophic interference.
Both evaluations on offline experiments and A/B test shows that our method outperforms all baselines considered.

\end{abstract}

\maketitle



\section{Introduction}

Click-through rate (CTR) prediction is a critical task in recommender systems and online advertising platforms. 
Most previous works mainly concentrates on better feature interaction\cite{wang2017deep,cheng2016wide,guo2017deepfm,liu2022position} and extensive user behavior modeling \cite{zhou2018deep,zhou2019deep,xiao2020deep,huang2021deep} based on deep neural networks trained with offline batch manner.

\begin{figure}[t]
  \centering
  \subfigure[Variation of CTR]{
  \includegraphics[width=0.24\textwidth]{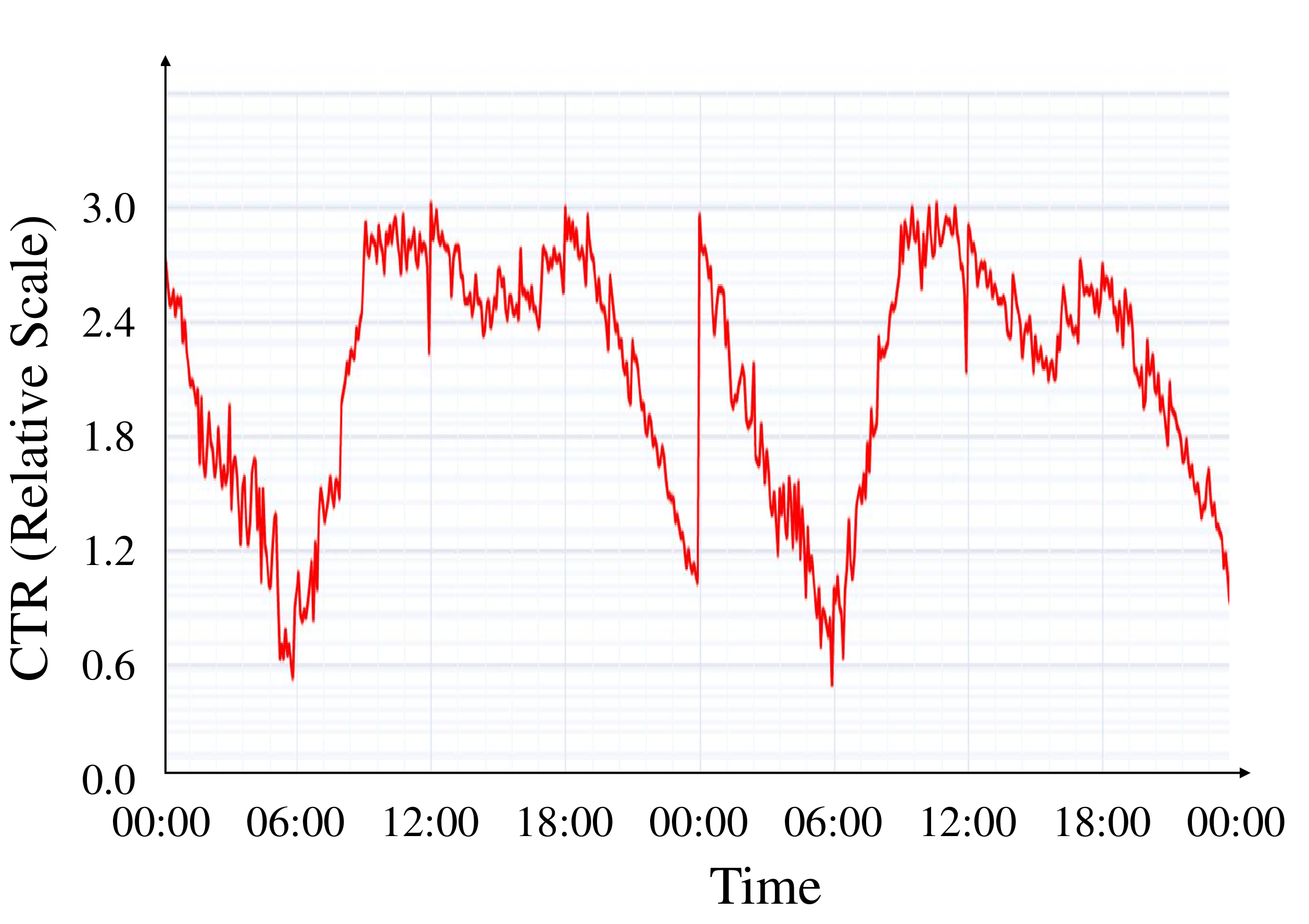}
  }
  \subfigure[8-hour versus 8-hour similarity]{
  \includegraphics[width=0.21\textwidth]{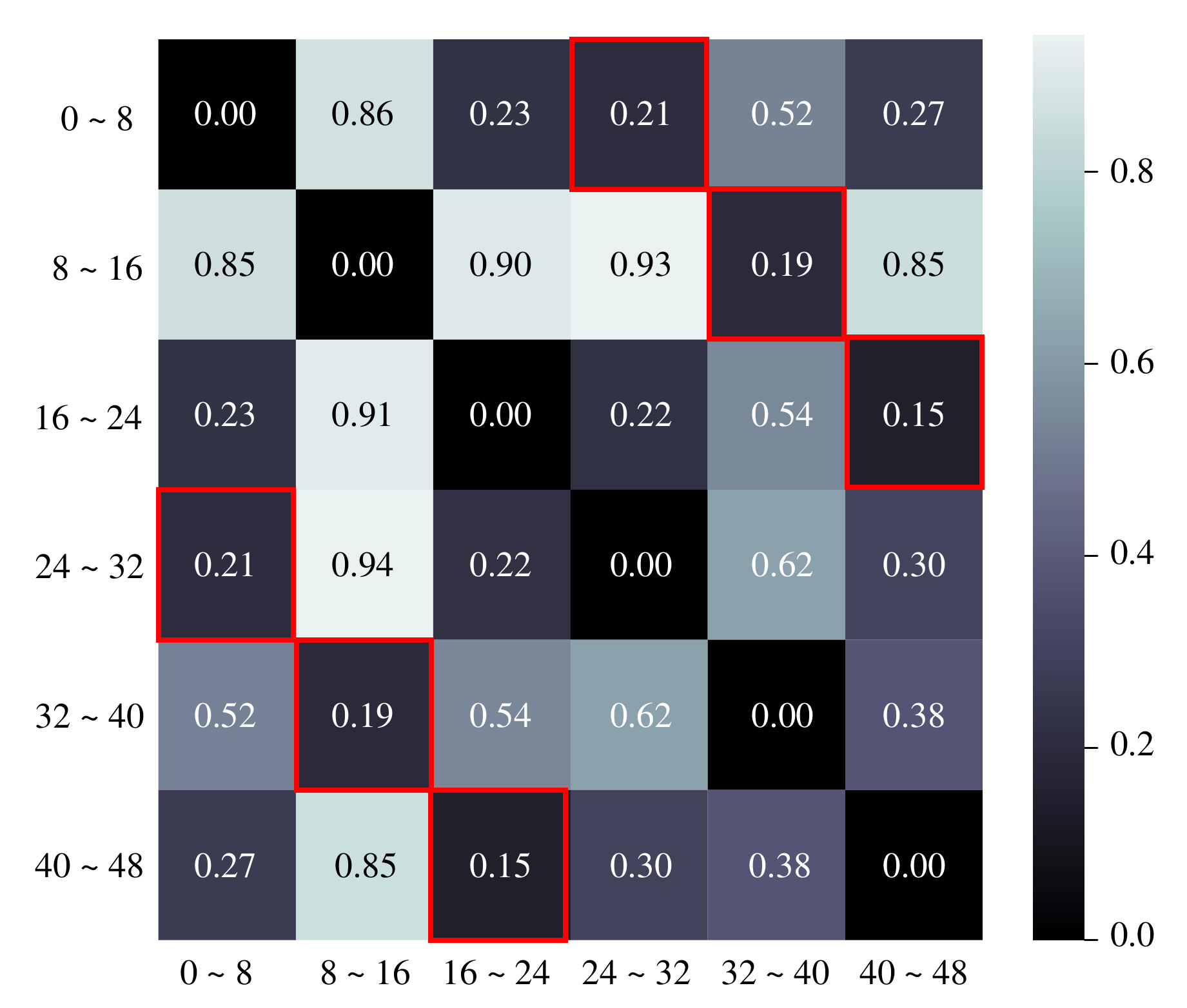}
  }
  \Description{Empirical observation.}
  \caption{Empirical observation of drifting and recurred patterns. (a) CTR in real production within a typical 2 days is shown in relative scale. (b) Impressed ad data of two days are grouped every eight hours, and the KL distance between groups of data is calculated and unified. smaller value means larger similarity.}
  \label{fig:drift}
\end{figure}

However, in real-world production systems such as online advertising platforms, CTR models are often trained with incremental learning framework to serve streaming data from massive users \cite{wang2020practical}. 
Frequently, the distribution of the streaming data drifts, manifested as patterns of rapid changes over time\cite{tsymbal2004problem,lu2018learning}, especially during the promotion period such as \textit{11.11} shopping Festival in Taobao and JD.com, and prime day in Amazon.
Furthermore, historical distribution may recur in the future \cite{recurrent2007Ramamurthy, lu2018learning}. 
Figure. \ref{fig:drift} provides an empirical observation of the drift phenomena in temporal distribution illustrated by CTR over time (a), and recurred patterns annotated by red box (b).

Vanilla incremental learning methods such as IncCTR proposed in \cite{wang2020practical} that keeping adapt to newest distribution observed may lead to catastrophic forgetting or catastrophic interference, 
since knowledge of the previously learned distribution may abruptly lost due to the adaptation to current distribution \cite{James2017, french1999catastrophic}. 
Furthermore, it's obviously inefficient to relearn distribution that has been occurred. 

Many effort has been devoted to overcoming the problem of catastrophic forgetting in neural network learning\cite{James2017, NIPS2017_f708f064, Liu_Yang_Wang_2021, pmlr-v139-chen21v, pmlr-v97-li19m}, which can basically be grouped into three classes\cite{9349197},
1) knowledge distillation\cite{Kang_2022_CVPR, 8107520, Dhar_2019_CVPR}: the influence of new knowledge on the network is reduced by introducing a regular term, thereby protecting old knowledge from being overwritten by new knowledge; 
2) replay\cite{Rebuffi_2017_CVPR, Lyu_Wang_Feng_Ye_Hu_Wang_2021}: prevent the model from forgetting old knowledge by retaining important old task data and reviewing the old tasks; 
3) parameter isolation\cite{aljundi2017expert, pmlr-v80-serra18a}: freeze the model parameters for storing old knowledge and adopt the new parameters for learning the new knowledge. 
Above methods have been widely used in many fields such as image classification and object detection\cite{8107520, Dhar_2019_CVPR, Rebuffi_2017_CVPR}. 
However, it's difficult to deploy these methods to large-scale CTR prediction for industrial scenarios due to memory constraints and diversity of data distributions.

Previous psychological research\cite{amer2022cluttered,murphy2022differential} show that human learners can strengthen their learning performance in memory tasks by filtering task-irrelevant information to avoid retroactive interference phenomena\cite{tulving1971retroactive} , i.e., new knowledge interfering with existing knowledge.
Similarly, neural network learners may also benefit from filtering out task-relevant information under drifting distributions based on performance metrics such as AUC, i.e., acquiring task-relevant information with high auc for memory consolidation.

In this paper, we propose a drift-aware incremental learning framework for CTR prediction, \textit{Always Strengthen Your Strengths} (ASYS), which explicitly detects drift of distribution with \textit{error rate-based} method
and integrates it into an ensemble learning framework to guide the update of different sub-learners. 
To better detect drift and preserve historical information, we improved on ADWIN \cite{bifet2007learning}, a widely used two-time window-based drift detection algorithm, to examine the adaptability of learners to incoming distribution with AUC as a performance metric.
ASYS aims to preserve differential information and prevent distribution corruption by multi-learner structures and modulating the update mechanism of learners, i.e., updates to strengthen the learner adapted to the current distribution and freezing updates of unadapted ones to avoid catastrophic interference.

Recent works with ensemble learning schemes learn differential knowledge between different ensembles and adopt a weighting scheme to aggregate final result for better performance in CTR prediction \cite{krawczyk2017ensemble,liu2022concept,yang2019adaptive}. However, these works adopted adaptive way to update models that do not explicitly detect drift of distributions and thereby is ineffective to address catastrophic forgetting.

This novel drift-aware framework greatly alleviate catastrophic forgetting when the distribution drifts over time and recur frequently.
To validate the effectiveness and efficiency of the proposed method, we conduct extensive experiments against various competing baselines for CTR prediction on a real-world industrial dataset and public dataset. 
Besides, we further demonstrate the performance of our \textit{ASYS} through a rigorous online A/B test in an online advertising system.
An intuitive qualitative analysis of AUC values over time is provided to illustrate the superiority of \textit{ASYS} in enhancing model performance and handling catastrophic forgetting.

The contributions of this paper are in four folds:
1). We proposed a novel drift-aware framework, \textit{ASYS}, for serving high-velocity user-generated streaming data. To the best of our knowledge, this is the \textit{first} attempt to incorporate explicit drift detection into ensemble learning to alleviate catastrophic forgetting in CTR prediction. 
2). We developed a novel drift detection method based on ADWIN, which is more suitable for us to use to address catastrophic forgetting. With explicit drift detection, our ASYS is with good generality and better interpretability.

3). We achieve significant improvements on a real-world industrial dataset over all incremental learning baselines. A rigorous A/B test further demonstrates the excellent performance over a highly optimized baseline models.  
4). We give a theoretical and qualitative analysis for better understanding.

\begin{figure}[!t]
  \includegraphics[width=0.5\textwidth]{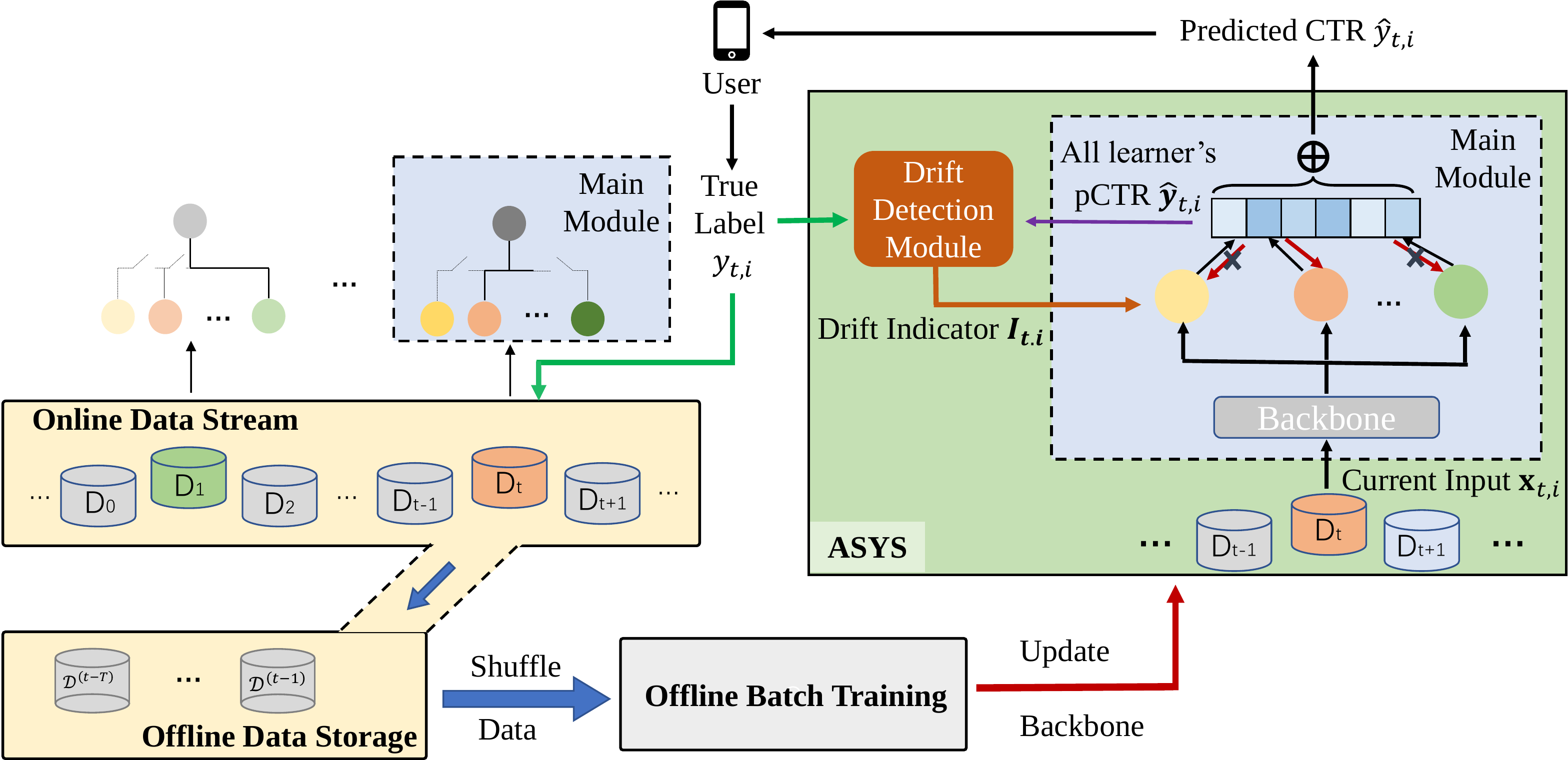}
  \caption{An overview of the proposed ASYS.}
  \Description{ASYS Module.}
  \label{fig:module}
\end{figure}

\section{Methodology}
\subsection{Problem Definition}
In this work, we consider the problem as learning on stream data for CTR prediction.
Given as stream of data $\{\mathcal{D}_t\}_{t=1}^{+\infty}$ where $\mathcal{D}_t = [\mathbf{x}_t, \mathbf{y}_t] = \left\{\mathbf{x}_{t, i}, y_{t, i}\right\}_{i=1}^{N_t}$ represents the training data at time step $t$.
$\mathbf{x}_{t, i}\in \mathbb{R}^f$ denotes the features of the $i$-th impressed advertisement at time step $t$ and $y_{t, i} \in \{0, 1\}$ denotes ground truth label that shows whether the user clicks the impressed advertisement or not.

As illustrated on the top right of Fig. \ref{fig:module}, the proposed \textit{ASYS} consists of two modules: the \textit{Drift Detection Module} detects whether drift occurs for every learner and determines which learners to update to preserve historical information, 
and the \textit{Main Module} uses input feature to predict the CTR and trains the model as directed by \textit{Drift Detection Module}. 
In this way, we can train each learner better and facilitate finer CTR prediction in \textit{Main Module}.

\subsection{Main Module}
In the main module, a backbone structure first converts the input sample $\mathbf{x}_{t, i}$ to embedding  $\mathbf{e}_{t, i}$ of the sample
\begin{equation}
  \mathbf{e}_{t, i} = Backbone(\mathbf{x}_{t, i}).
\end{equation}

The backbone structure can be suitable DNNs such as deep interest network (DIN) \cite{zhou2018deep} or other networks.
Then, the extracted embedding $\mathbf{e}_{t, i}$ is fed into $m$ learners and they output their own predicted CTR (pCTR).
The structure of each learner is an MLP network followed by a sigmoid activation function.

At inference time, we aggregate all pCTRs to obtain predicted CTR of the sample $\mathbf{x}_{t, i}$, but when training the model, we aggregate these pCTRs under the instruction of drift detection module,

\begin{equation}
\begin{aligned}
  & \hat{y}_{t, i}^{infer} = \mathbf{w}_{t} \cdot \hat{\mathbf{y}}_{t, i} = \sum_{k=1}^{m} w^{(k)}_{t} \hat{y}_{t, i}^{(k)}, \\
  & \hat{y}_{t, i}^{train} = Normal(\mathbf{w}_{t} \odot \mathbf{I}_{t})\cdot \hat{\mathbf{y}}_{t, i} = \sum_{k=1}^{m} \frac{w^{(k)}_{t} I^{(k)}_{t}}{||\mathbf{w}_{t} \odot \mathbf{I}_{t}||_1} \hat{y}_{t, i}^{(k)},
  \label{eqn:y}
\end{aligned}
\end{equation}
where $\mathbf{w}_t = [w^{(1)}_{t}, \cdots, w^{(m)}_{t}]^T \in \mathbb{R}^{m}$ is the aggregation weights at time step $t$, $\mathbf{I}_{t} \in \{0, 1\}^{m}$ is the indicator given by drift detection module at time step $t$, and $\hat{\mathbf{y}}_{t, i} = [\hat{y}_{t, i}^{(1)}, \cdots \hat{y}_{t, i}^{(m)}]^T \in [0, 1]^{m}$ is the concatenation of the learners' pCTR, where $\hat{y}_{t, i}^{(k)} = Learner^{(k)}(\mathbf{e}_{t, i})$. The symbol $\odot$ in Eqn. (\ref{eqn:y}) means element-wise multiplication, and the $Normal$ operator aims to force the sum of the aggregation weight of all learners to be 1. Then we use $\hat{y}_{t, i}^{train}$ to compute loss and do backpropagation.
The rule to update aggregation weights $\mathbf{w}_t$ at time step $t$ can be IncCTR\cite{wang2020practical}, MoE\cite{jacobs1991adaptive}, AdaMoE\cite{liu2022concept} or others.

\subsection{Drift Detection Module}

To examine the fitness of the learner to incoming distribution, we build on ADWIN \cite{bifet2007learning} for distribution drift detection, a widely used error rate based method. 
Unlike ADWIN which develops for drift detection and adaptation to new distribution, we perform drift detection to protect the information of the historical distribution from being corrupted in case of a drifted new distribution occurs.

Based on the purpose of assessing model validity and thus detecting drift, we use AUC as a metric for learners' performance. We keep a sliding window of fixed length $L$ for each learner to store the historical information and put the first $L$ AUCs sequentially into the corresponding sliding window as initial historical information. Every time a new AUC arrives, we assume that the AUC will significantly decrease if the current distribution drifts, thus we test whether the observed average of the more recent AUCs is significantly lower than that of the older AUCs.
\begin{algorithm}[t]
\SetAlgoLined
\small
\footnotesize
\KwIn{Data $\mathcal{D}_t$, Length $L$ \quad \textbf{Output: }Predicted CTR $\hat{y}_{t, i}^{infer}$}

  $\hat{\mathbf{y}}_{t, i}, \mathbf{w}_t \leftarrow \text{MainModule}(\mathbf{x}_{t, i})$;\\

  {\nonl\footnotesize\texttt{// Inference: predict CTR}}\;
  
  $\hat{y}_{t, i}^{infer} \leftarrow \mathbf{w}_{t} \cdot \hat{\mathbf{y}}_{t, i}$;
  \hspace{0.5em}{\footnotesize\texttt{// Aggregate pCTR of all Learners}}\;
  
  {\nonl\footnotesize\texttt{// Train: update the model}}\;
  
  \nonl \textbf{for} each learner $k$ \textbf{do} \\
  
  \quad $auc_t^k \leftarrow \text{AUC}(\hat{y}_{t, i}^{(k)}, \mathbf{y}_t)$;
  \hspace{0.5em}{\footnotesize\texttt{// Compute auc}}\;
  
  \quad $\tilde{W}^k \leftarrow W_{t-1}^k \cup \{auc_t^k\}$;
  \hspace{0.5em}{\footnotesize\texttt{// Extend sliding window}}\;

  \quad $\epsilon_t^k \leftarrow \max_{L_{cut}}(\epsilon_{partion}-\epsilon_{cut})$ ;
  \hspace{0.5em}{\footnotesize\texttt{// Detection}};\\

  \nonl \quad \textbf{if not drift}  
  \hspace{0.5em}{\footnotesize\texttt{// Update sliding window and indicator}}\;
  \quad \quad $W_t^k, I_t^{(k)} \leftarrow \tilde{W}^k \backslash W_{1, t-1}^k,  1$; \\
  \nonl \quad \textbf{else}\\
  \quad \quad $W_t^k, I_t^{(k)} \leftarrow W_{t-1}^k, 0$; \\
  \nonl \quad \textbf{end if}\\
  \nonl \textbf{end for}\\
  $\hat{y}_{t, i}^{train} \leftarrow Normal(\mathbf{w}_{t} \odot \mathbf{I}_{t})\cdot \hat{\mathbf{y}}_{t, i}$;
  \hspace{0.5em}{\footnotesize\texttt{// Aggregate pCTR}}\;
  Use $\hat{y}_{t, i}^{train}$ to compute loss and do backpropagation.
  \caption{ASYS for CTR Prediction.}
\end{algorithm}

For $k$th-learner at time step $t$, denote the AUC as $auc_t^k$ and the sliding window as $W_t^k = \{W_{l, t}^k\}_{l=1}^L$. Denote $\tilde{W}^k = \{\tilde{W}_l^k\}_{l=1}^{L+1}$ as the temporary extended sliding window, where $\tilde{W}_l^k=W_{l, t}^k$ for $l \leq L$, and $\tilde{W}_{L+1}^k=auc_{t+1}^k$. Like ADWIN\cite{bifet2007learning}, we get a partion of $\tilde{W}^k$, and the cut point is $L_{cut}$. Specifically, we test whether $\epsilon_{partion}$ is lower than $\epsilon_{cut}$, where

\begin{equation}
\begin{aligned}
  & \epsilon_{partion} = \frac{1}{L-L_{cut}+1}\sum_{l=L_{cut}+1}^{L+1}\tilde{W}_l^k - \frac{1}{L_{cut}}\sum_{l=1}^{L_{cut}}\tilde{W}_l^k, \\
  & \epsilon_{cut} = -(\max \tilde{W}_l^k - \min \tilde{W}_l^k)\sqrt{\frac{1}{2n_0}\cdot ln\frac{2}{\delta^{\prime}}}, \delta^{\prime} = \frac{\delta}{L+1},
  \label{eqn:detect}
\end{aligned}
\end{equation}
$n_0$ is the harmonic mean of $L_{cut}$ and $L+1-L_{cut}$, and $\delta$ is confidence level. The formula of $\epsilon_{cut}$ is different from that of \cite{bifet2007learning} because we use one-sized test instead of two-sized test. Theorem 3.1 in \cite{bifet2007learning} ensures that the probability of such a detection making a Type I error does not exceed $\delta$. Inspired by exponential histogram technology\cite{bifet2007learning, Datar2016, doi:10.1137/S0097539701398363} which guarantees $(1+\epsilon)$-multiplicative approximation, we iterate $L_{cut}$ in $S := \{L+1-2^z|z\in\mathbb{Z}_{\geq 0}, L+1-2^z>0\}$ and let $\epsilon_t^k = \min_{L_{cut}}(\epsilon_{partion}-\epsilon_{cut})$. Two situations that may occur:

\begin{itemize}
    \item $\max_k\epsilon_t^k>0$, which means not all learners' distribution drifts. 
    In this case, we freeze learners that do not satisfy $\epsilon_t^k>0$, i.e., $I^{(k)}_{t}=\mathbb{I}_{\epsilon_t^k>0}$.
    Besides, we update the sliding window for learners not frozen, $W_{t+1}^k = \tilde{W}^k \backslash W_{1, t}^k$, for other learners $W_{t+1}^k=W_t^k$. We prevent the current information from corrupting their preserved historical information by freezing some learners and preventing the backpropagation of their parameter gradients. 
    \item $\max_k\epsilon_t^k<0$, which means the current information drifts for all learners. Denote $k_0 = \arg\max_k \epsilon_t^k$, at this time we freeze all except $k_0$th-learner to make a trade-off between retaining historical information and learning current information.
\end{itemize}

We briefly illustrate the effect of the length $L$ of the sliding window. If $L$ is too small, the fluctuation will be more likely detected as distribution drift, and the result will be unstable. If $L$ is too large, the smaller AUC in the past is still retained when the larger AUC enters the sliding window. According to Eqn. (\ref{eqn:detect}), $\epsilon_{cut}$ will become smaller, cause drift to be more difficult to detect and the drift detection module will gradually fail.

\subsection{Theoretical Analysis of ASYS}
\begin{myTheo} 
    \label{thm:bias_variance_cr}
    \begin{sloppypar}
    (\textbf{Upper Bound for Estimation Error})\label{prop_cr} 
    Suppose $P(x, y)$ and $P^{\prime}(x, y)$ are two joint distributions of $x$ and $y$. For any Lipschitz continuous $f$ of Lipschitz constant $L$ and $y \in \{0, 1\}$, $\mathbf{x}_i\sim P(\cdot|y)$ and $\mathbf{x}_j\sim P^{\prime}(\cdot|y)$, denote $\mathbb{E}_{(\mathbf{x}_i, y)\sim P^{\prime}}||f(\mathbf{x}_i)-y||=\epsilon_{P^{\prime}}$, we have
    	\begin{equation}
    		\label{bound_cr}
    		\mathbb{E}_{\mathbf{x}_i\sim P(\cdot|y)}||f(\mathbf{x}_i)-y|| \leq \mathbb{E}_{\mathbf{x}_i, \mathbf{x}_j\sim P(\cdot|y), P^{\prime}(\cdot|y)}L||\mathbf{x}_i-\mathbf{x}_j||+\epsilon_{P^{\prime}}, 
    	\end{equation}
    \end{sloppypar}
\end{myTheo}

\textbf{Theoretical Understanding of ASYS. } The above theorem can be easily proved by norm inequality and definition of Lipschitz continuity. Suppose sequential finite data $\mathcal{D}$, $\mathcal{D}^{\prime}$ is sampled from $P(x, y)$, $P^{\prime}(x, y)$ respectively, and we get function $f$ by training on $\mathcal{D}$ and $\mathcal{D}^{\prime}$. 
Drift can be triggered by 3 source from \cite{lu2018learning}: $P(x) \neq P^{\prime}(x)$ (source 1), $P(y|x) \neq P^{\prime}(y|x)$ (source 2), both (source 3). 
When $P^{\prime}$ drift relative to $P$, $P^{\prime}(x|y)$ drift relative to $P(x|y)$ no matter which kind of drift appears since the Bayes formula $P(x|y) \propto P(x)P(y|x)$, thus term $\mathbb{E}_{\mathbf{x}_i, \mathbf{x}_j\sim P(\cdot|y), P^{\prime}(\cdot|y)}L||\mathbf{x}_i-\mathbf{x}_j||$ will be large, the upper bound will be large after training on $\mathcal{D}^{\prime}$. 
Our ASYS freezes those learners whose indicator is $0$, thus keep the upper bound of each learner at a low level. Each model therefore corresponds to a class of distributions with a lower upper bound on the data sampled from such distributions, which alleviate the catastrophic forgetting.

Besides, the variance of the predicted CTR is the weighted sum of the variance of all pCTRs. When there exists some learners with lower upper bound for recurring distribution, the variance of the predicted CTR will also be lower, improving model's performance.

\section{Experiments}

\begin{table}[t]
  \caption{The Overall AUC on Avazu and Industrial dataset over 3-Run, std$\approx10^{-4}$. \#(Lea.) denotes number of learners.}
  \label{tab:benchmark-res}
  \small
\resizebox{\columnwidth}{!}{ 
\begin{tabular}{llc|cccccc}
\hline
\multicolumn{2}{c}{\multirow{2}{*}{}}           & \multicolumn{1}{l|}{\multirow{2}{*}{\#Lea.}} & \multirow{2}{*}{IncCTR} & \multicolumn{1}{l}{\multirow{2}{*}{\begin{tabular}[c]{@{}l@{}}IncCTR\\ +ASYS\end{tabular}}} & \multirow{2}{*}{MoE} & \multicolumn{1}{l}{\multirow{2}{*}{\begin{tabular}[c]{@{}l@{}}MoE\\ +ASYS\end{tabular}}} & \multirow{2}{*}{AdaMoE} & \multicolumn{1}{l}{\multirow{2}{*}{\begin{tabular}[c]{@{}l@{}}AdaMoE\\ +ASYS\end{tabular}}} \\
\multicolumn{2}{c}{}                            & \multicolumn{1}{l|}{}                            &                         & \multicolumn{1}{l}{}                                                                        &                      & \multicolumn{1}{l}{}                                                                     &                         & \multicolumn{1}{l}{}                                                                        \\ \hline
\multicolumn{2}{l}{\multirow{4}{*}{\rotatebox{90}{\textbf{Avazu}}}}      & 3                                                & 0.7603                  & \textbf{0.7617}                                                                                     & 0.7600               & \textbf{0.7619}                                                                                   & 0.7618                  & \textbf{0.7634}                                                                                      \\
\multicolumn{2}{l}{}                            & 6                                                & 0.7605                  & \textbf{0.7621}                                                                                      & 0.7605               & \textbf{0.7623}                                                                                   & 0.7621                  & \textbf{0.7636}                                                                                      \\
\multicolumn{2}{l}{}                            & 9                                                & 0.7595                  & \textbf{0.7622}                                                                                      & 0.7607               & \textbf{0.7620}                                                                                  & 0.7624                  & \textbf{0.7632}                                                                                      \\
\multicolumn{2}{l}{}                            & 12                                               & 0.7578                  & \textbf{0.7621}                                                                                      & 0.7603               & \textbf{0.7617}                                                                                   & 0.7620                  & \textbf{0.7630}                                                                                      \\ \hline
\multicolumn{2}{l}{\multirow{4}{*}{\rotatebox{90}{\textbf{Industrial}}}} & 3                                                & 0.7631                  & \textbf{0.7663}                                                                                     & 0.7627               & \textbf{0.7658}                                                                                   & 0.7635                  & \textbf{0.7667}                                                                                      \\
\multicolumn{2}{l}{}                            & 6                                                & 0.7634                  & \textbf{0.7663}                                                                                      & 0.7633               & \textbf{0.7660}                                                                                   & 0.7644                  & \textbf{0.7669}                                                                                      \\
\multicolumn{2}{l}{}                            & 9                                                & 0.7638                  & \textbf{0.7666}                                                                                      & 0.7632               & \textbf{0.7661}                                                                                   & 0.7645                  & \textbf{0.7670}                                                                                      \\
\multicolumn{2}{l}{}                            & 12                                               & 0.7634                  & \textbf{0.7658}                                                                                      & 0.7631               & \textbf{0.7657}                                                                                   & 0.7634                  & \textbf{0.7665}                                                                                      \\ \hline
\end{tabular}}
\end{table}

\subsection{Offline Experiment}
\textbf{Datasets.} 1). The industrial dataset is constructed from the user logs of real production systems.
The dataset contains about 6 billion ad impression records within three days, sorted in chronological order. 
Then it will be divided into data chunks $\mathcal{D}_t$ , each containing 2048 samples.
2). The public dataset follows settings of AdaMoE  \cite{liu2022concept}
which reorganized the Avazu\footnote{\url{https://www.kaggle.com/c/avazu-ctr-prediction/data}} dataset chronologically.

\begin{figure}[t]
  \centering
  \subfigure[AUC/10Min of different models]{
  \includegraphics[width=0.236\textwidth]{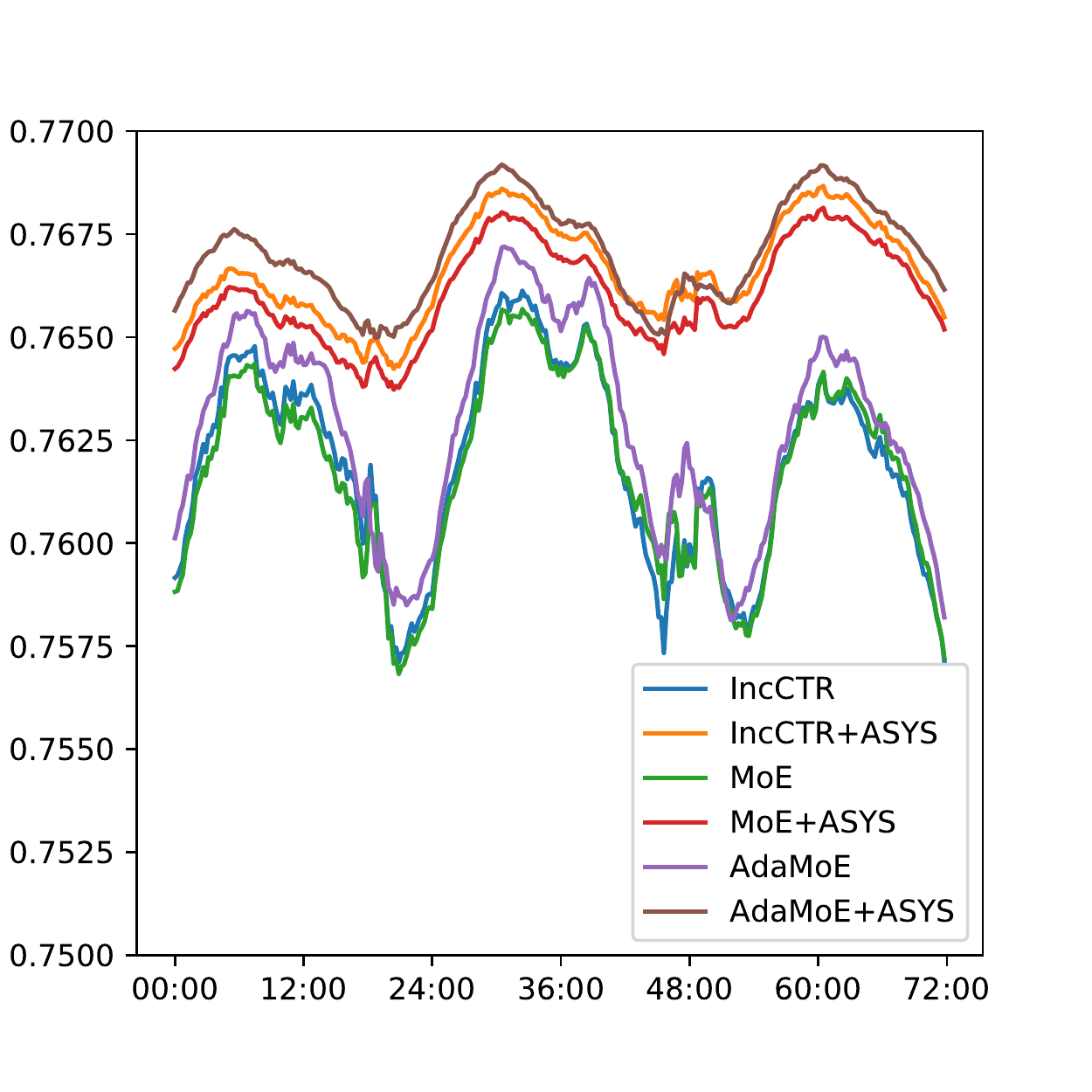}
  \label{fig:auc-models}
  }\hspace{-10pt}
  \subfigure[AUC/10Min of different $\delta$]{
  \includegraphics[width=0.236\textwidth]{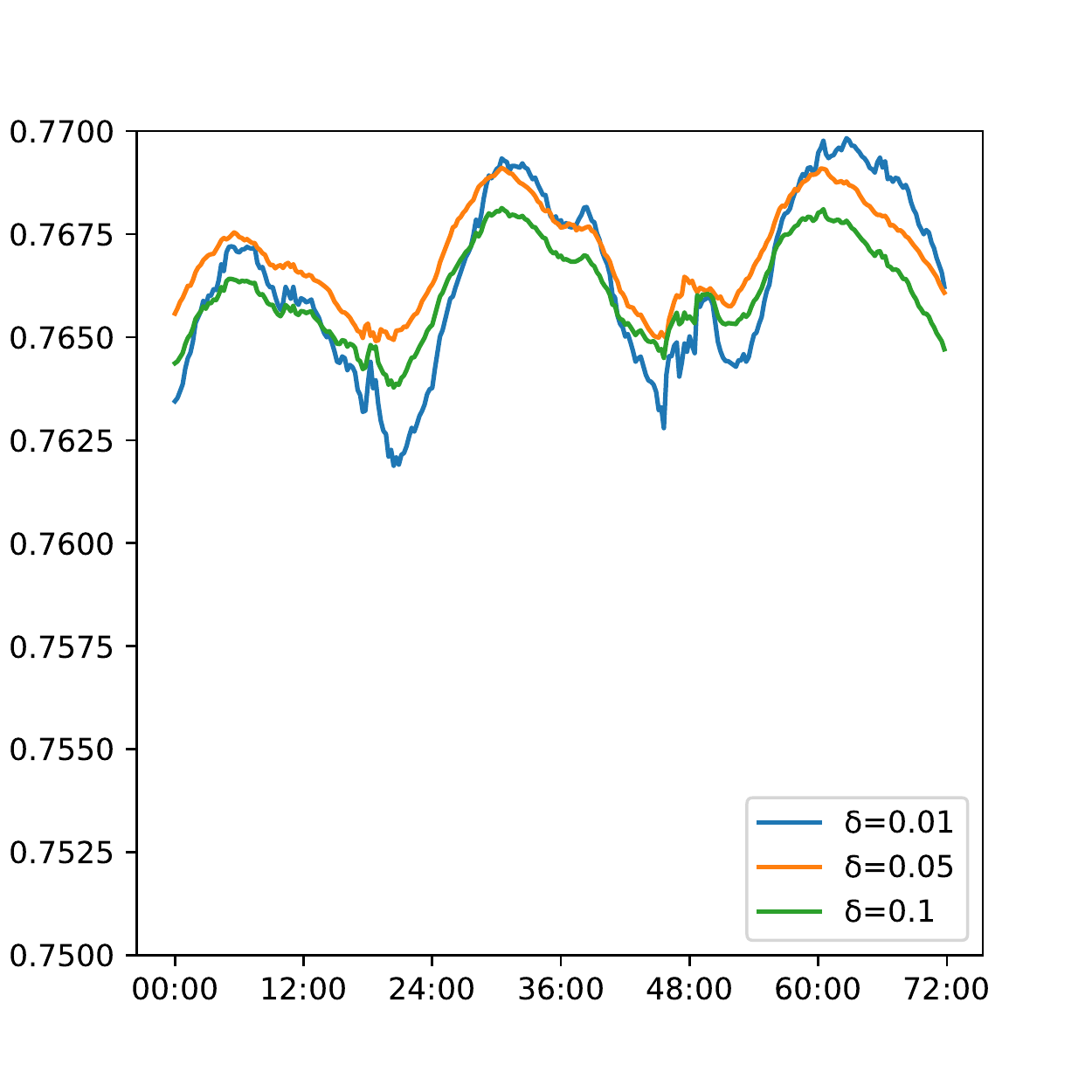}
  \label{fig:auc-delta}
  }
  \Description{Drift adaptation results.}
  \caption{AUC/10Min of different (a) models and (b) $\delta$.}
  \label{fig:auc-change-results}
\end{figure}

\textbf{Baseline.} We apply ASYS with the following baseline models.

$\bullet$ \textit{IncCTR}: The pCTR is simply the average of all learners' output.

$\bullet$ \textit{MoE}: The outputs of learners are aggregated by a gate network.

$\bullet$ \textit{AdaMoE}: The update of aggregation module is decoupled from regular MoE following settings in \cite{liu2022concept}.

\textbf{Settings.} We perform offline batch training by shuffling the offline data in storage, and then update the backbone as shown in Fig. \ref{fig:module}. An ensemble of DIN \cite{zhou2018deep} and DCN\cite{wang2017deep} is used as the backbone with output dimension of 1024.
Each learner (expert) is a 2-layer-MLP with [512, 256] hidden units with a ReLU activation function, followed by a linear layer with the sigmoid function to obtain the predicted CTR. 
Adam optimizer is used for all models. For hyper-parameters of considered baselines, we follow the same setting with \cite{liu2022concept}.

\textbf{Evaluation Metrics.} 
We adopt AUC (Area Under ROC Curve) as the evaluation metric.
In our setup, data is streamed and used for training only once.
For data chunk $\mathcal{D}_t$ at time step $t$, its first 80\% of data are used as training samples $\mathcal{D}_t^{train}$, and the rest 20\% as testing samples $\mathcal{D}_t^{test}$.
At each time step $t$, $\mathcal{D}_t^{train}$ is used to train the model first.
Then, the trained model is used to $\mathcal{D}_t^{test}$ to obtain pCTRs.
The above train-predict process continues until the entire data stream is consumed.
To compute \textit{overall AUC}, we use the pCTRs and labels of all test samples.
Consistent with \cite{liu2022concept},we further collect the stored pCTR and ground truth label of test samples every 10 minutes to calculate \textit{AUC every 10 minutes (AUC/10Min)} to better study the model performance to stream data.

\textbf{Quantitative Results.} The confidence level is set to $\delta=0.05$. As shown in Table. \ref{tab:benchmark-res},
ASYS overwhelms all baseline models with \#Lea. varying from 3 to 12 in both industrial dataset and the public Avazu dataset.
When \#Lea. is nine, ASYS improves the overall AUC by 0.34\% on average over all baseline methods in industrial dataset which is significant in CTR prediction \cite{wang2017deep,cheng2016wide,guo2017deepfm}. Similar results are observed over public dataset with 0.18\%-0.35\% improvements.

For ASYS, we obtain the best result when \#Lea. is nine in industrial dataset and six in in Avazu dataset. We can see that too many learners can result in the decrease of overall AUC. The reason may be that too many parameters cause underfitting.

\textbf{Qualitative Results.} To further study how the models handle catastrophic forgetting in streaming data, we plot the AUC/10Min in Fig. \ref{fig:auc-models}.
The \#Lea. is set to nine. Similar results can be observed with other \#Lea.
We can observe that the distribution of time periods such as 00:00 and 12:00 of each day is similar everyday, reflecting the periodicity of the distribution. 
For all baselines, the AUCs drops significantly at 00:00, indicating that the model catastrophically forgets the distribution around 00:00 after learning the distribution at other times. After adding our ASYS to each baseline model, the drop is greatly alleviated and AUC is greatly improved.

\textbf{Influence of confidence level in drift detection module.} 
We set the baseline model as AdaMoE because of its best performance and investigate the impact of $\delta$, as shown in Fig. \ref{fig:auc-delta}. We find that AUC under $\delta=0.05$ is better than $\delta=0.01$ and $\delta=0.1$, and the drop of AUC for $\delta=0.01$ is indeed larger than $\delta=0.05$ and $\delta=0.1$. 
The reason might be that the smaller $\delta$ is, the looser the detection is and the degree of overcoming catastrophic forgetting is not enough. While when $\delta$ is large, the probability of making Type I error is larger that is easier to misdetect leading to insufficient training. We empirically find $\delta=0.05$ achieves a better tradeoff between strict and lenient detection of ASYS, and yields the best overall AUC.

\subsection{Online A/B Test}

To conduct online A/B test, we deploy the proposed ASYS model in the online advertising system.
The online A/B test lasts for 7 days (from 2022-Aug-25 to 2022-Sep-1).
Compared to a highly optimized baseline, our ASYS contributes to 2.62\% CTR and 3.56\% eCPM (Effective Cost Per Mille) gain.

\section{Conclusion}
In this paper, we introduce a novel incremental learning framework, ASYS, to address the catastrophic forgetting in CTR prediction.
The experiments show that our method overwhelms all other incremental learning methods on a real-world production dataset and public dataset.
The online A/B test results further demonstrate the effectiveness of the proposed method.
A theoretical analysis and qualitative results are provided to illustrate the superior performance of ASYS on data streams. 
To the best of our knowledge, this is the first work to explicitly detect drift to handle catastrophic forgetting in the CTR prediction.

\balance
\bibliographystyle{ACM-Reference-Format}
\bibliography{sample-base}

\end{document}